\documentclass[twocolumn,superscriptaddress,pre]{revtex4}

\usepackage{color}
\usepackage{amsmath}      
\usepackage{amssymb} 
\usepackage{amsfonts}   
\usepackage{graphicx}
\usepackage{pgfplots}
\usepackage{bm}
\usepackage{ifthen}
\usepackage{lipsum}

\usepackage{bbm}
\usepackage{setspace}
\usepackage{float}
  
\newcommand{\bea}{\begin{eqnarray}}
\newcommand{\eea}{\end{eqnarray}}
\newcommand{\beq}{\begin{equation}}
\newcommand{\eeq}{\end{equation}}

\newcommand{\Anonym}[1]{{\color[rgb]{0,0,0}}}

\newcommand{\UF}[1]{{\color[rgb]{0.0,0.0,0.0}#1}}
\newcommand{\TA}[1]{{\color[rgb]{0.0,0.0,0.}#1}}

\begin{document}

\title{Maximum entropy models reveal the excitatory and inhibitory correlation structures in cortical neuronal activity}

\author{Trang-Anh Nghiem}
\affiliation{Laboratory of Computational Neuroscience, Unit\'e de Neurosciences, Information et
Complexit\'e, CNRS, Gif-Sur-Yvette, France.}
\author{Bartosz Telenczuk}
\affiliation{Laboratory of Computational Neuroscience, Unit\'e de Neurosciences, Information et
Complexit\'e, CNRS, Gif-Sur-Yvette, France.}
\author{Olivier Marre}
\affiliation{Sorbonne Universit\'e, INSERM, CNRS, Institut de la Vision, 17 rue Moreau, 75012 Paris, France}
\author{Alain Destexhe}
\thanks{These authors contributed equally.}
\affiliation{Laboratory of Computational Neuroscience, Unit\'e de Neurosciences, Information et
Complexit\'e, CNRS, Gif-Sur-Yvette, France.}
\author{Ulisse Ferrari}
\thanks{These authors contributed equally.}
\affiliation{Sorbonne Universit\'e, INSERM, CNRS, Institut de la Vision, 17 rue Moreau, 75012 Paris, France}
\affiliation{Correspondence should be sent to \url{ulisse.ferrari@gmail.com}.}

\begin{abstract}
\TA{Maximum Entropy models can be inferred from large data-sets to uncover how collective dynamics emerge from local interactions.
Here, such models are employed to investigate neurons recorded by multielectrode arrays in the human and monkey cortex. 
Taking advantage of the separation of excitatory and inhibitory neuron types, we construct a model including this distinction.
This approach allows to shed light upon differences between excitatory and inhibitory activity across different brain states such as wakefulness and deep sleep, in agreement with previous findings. 
Additionally, Maximum Entropy models can also unveil novel features of neuronal interactions, which are found to be dominated by pairwise interactions during wakefulness, but are population-wide during deep sleep. 
In particular, inhibitory neurons are observed to be strongly tuned to the inhibitory population.
Overall, we demonstrate Maximum Entropy models can be useful to analyze data-sets with classified neuron types, and to reveal the respective roles of excitatory and inhibitory neurons in organizing coherent dynamics in the cerebral cortex. }\\

\end{abstract}

\maketitle

\bigskip

\section{INTRODUCTION}
To analyze a complex system, one is interested in finding 
a model able to explain the most about empirical data, with the fewest forms of
interactions involved. Such a model should reproduce the statistics
observed in the data, while making the least possible number of assumptions on
the structure and parameters of the system. In other terms, one needs
the simplest, most generic model that generates statistics
matching the empirical values - this implies \textit{maximising
entropy} in the system, with constraints
imposed by the empirical statistics \citep{Jaynes82}.

In a seminal paper \citep{Schneidman06}, a framework equivalent to
the Ising model in statistical physics was used to analyze the collective
behavior of neurons. This approach was based on the assumption that
pairwise interactions between neurons can account for the collective
activity of the neural population. Indeed, it was shown for experimental
data, from the retina and cerebral cortex, that this approach can
predict higher order statistics, including the probability distribution of 
the whole population's spiking activity. 
Even though the empirical pairwise
correlations were very weak, the model performed significantly better
than a model reproducing only the firing rates without considering
correlations. The Ising model was subsequently demonstrated to
efficiently reproduce the data better than models with smaller entropy
\citep{Ferrari17}, as well as to analyse neural recordings in a
variety of brain regions in different animals, ranging from the
salamander retina \citep{Schneidman06,Tkacik14} to the cerebral cortex
of mice \citep{Hamilton13}, rats \citep{Tavoni17}, and cats
\citep{Marre09}.

A complementary approach was recently introduced \citep{Okun15},
aiming at reproducing the correlation between single neuron
activity and whole-population dynamics in the mouse and monkey visual
cortex. \UF{This approach has then been generalized \citep{Gardella16}} to model
the neurons' full profile of dependency with the population activity,
and applied the model to the salamander retina.
\UF{Later work \citep{Odonnell17} further investigates the properties of these models with neuron-to-population couplings.}

Recent advances in experimental methods have allowed the recording of the
spiking activity of up to a hundred neurons throughout hours of
wakefulness and sleep, for instance using multi-electrode arrays, also
known as Utah arrays. Inspection of neurons' spike waveforms and their
cross-correlograms with other neurons made the discrimination of
excitatory (E) and inhibitory (I) neuron types possible
\citep{Peyrache12, Dehghani16}. Such data-sets therefore provide a further step in
the probing of the system, due to the unprecedented availability
of the simultaneously recorded dynamics of E and I neurons. 

In the present paper, we apply Maximum Entropy \UF{(MaxEnt)} models to
analyze human and monkey Utah array recordings.
We investigate in which way such models may
describe the two recorded (E, I) populations. 
\UF{As a proof of concept, we demonstrate how this approach can be applied to investigate 
excitatory and inhibitory neural activity across the brain states of wakefulness and deep sleep.}

\begin{figure*}[t!]
\begin{center}
\includegraphics[clip=true,keepaspectratio,width=1.95\columnwidth]{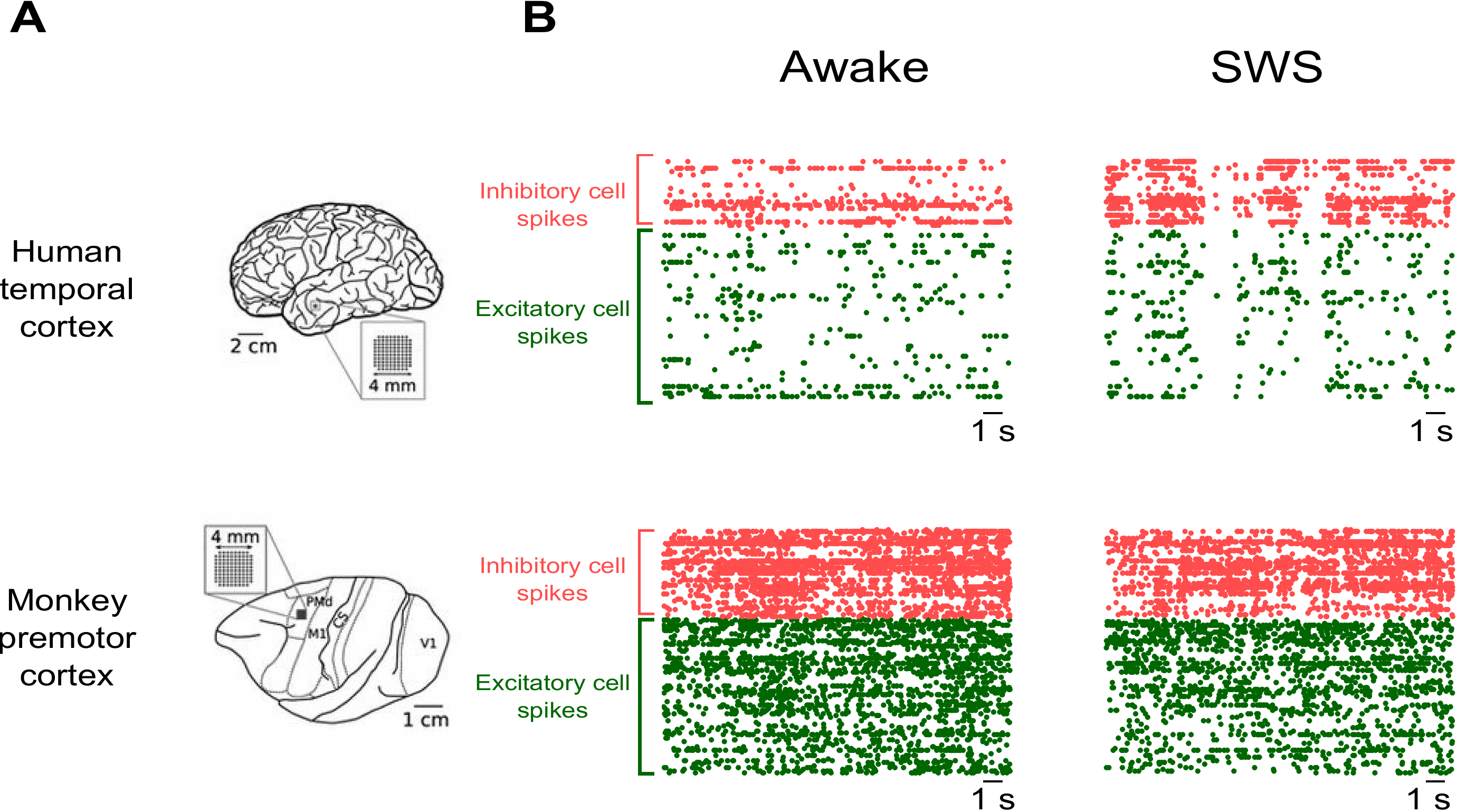}
\caption{\textbf{Multi-electrode (Utah) array recordings.} 
\textbf{A}) Utah array position in human temporal cortex (top) and monkey prefrontal cortex (bottom). Figure adapted from \citep{Telenczuk17}. 
\textbf{B}) Raster plots of spikes recorded for human (top) and monkey (bottom) in wakefulness (left) and SWS (right). \TA{Neurons are ordered to separate excitatory (E) from inhibitory (I) cells}
}
\label{fig:raw_data}
\end{center}
\end{figure*}

\begin{figure*}[t!]
\includegraphics[clip=true,keepaspectratio,width=1.995\columnwidth]{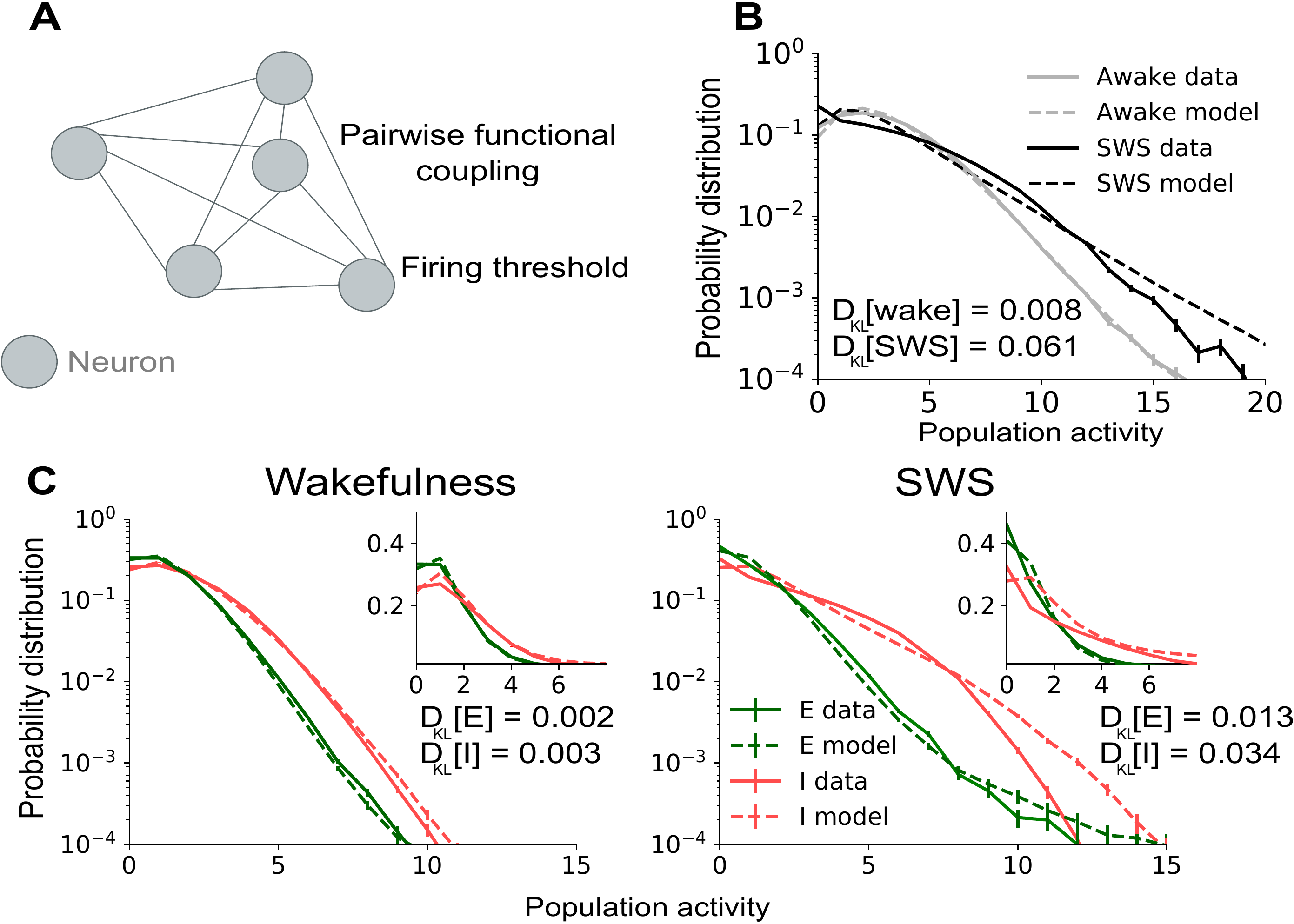}
\caption{
\textbf{Pairwise Ising model fails to predict SWS synchronous activity, especially for inhibitory neurons.}
{\bf A}) Model schematic diagram. Model parameters are each neuron's bias toward firing, and symmetric functional couplings between each pair of neurons. 
{\bf B}) \TA{Empirical and predicted probability distributions of the population activity $K = \sum_i \sigma_i$ for the neuronal population. 
The Ising model more successfully captures the population statistics during wakefulness than SWS, especially for medium and large $K$ values. 
{\bf C}) Empirical and predicted population activities for E (lower curves, in green/dark grey) and I (upper curves, in red/light grey) neurons. }The model particularly fails at reproducing the statistics of I population activity. These results are consistent with the presence of transients of high activity and strong synchrony between I neurons during SWS. \UF{Insets show an enlarged view on the region of low population activity, region within which the system spends the vast majority of the time (on a linear scale)}. 
}
\label{fig:Ising}
\end{figure*}

\begin{figure*}[t!]
\includegraphics[clip=true,keepaspectratio,width=1.99\columnwidth]{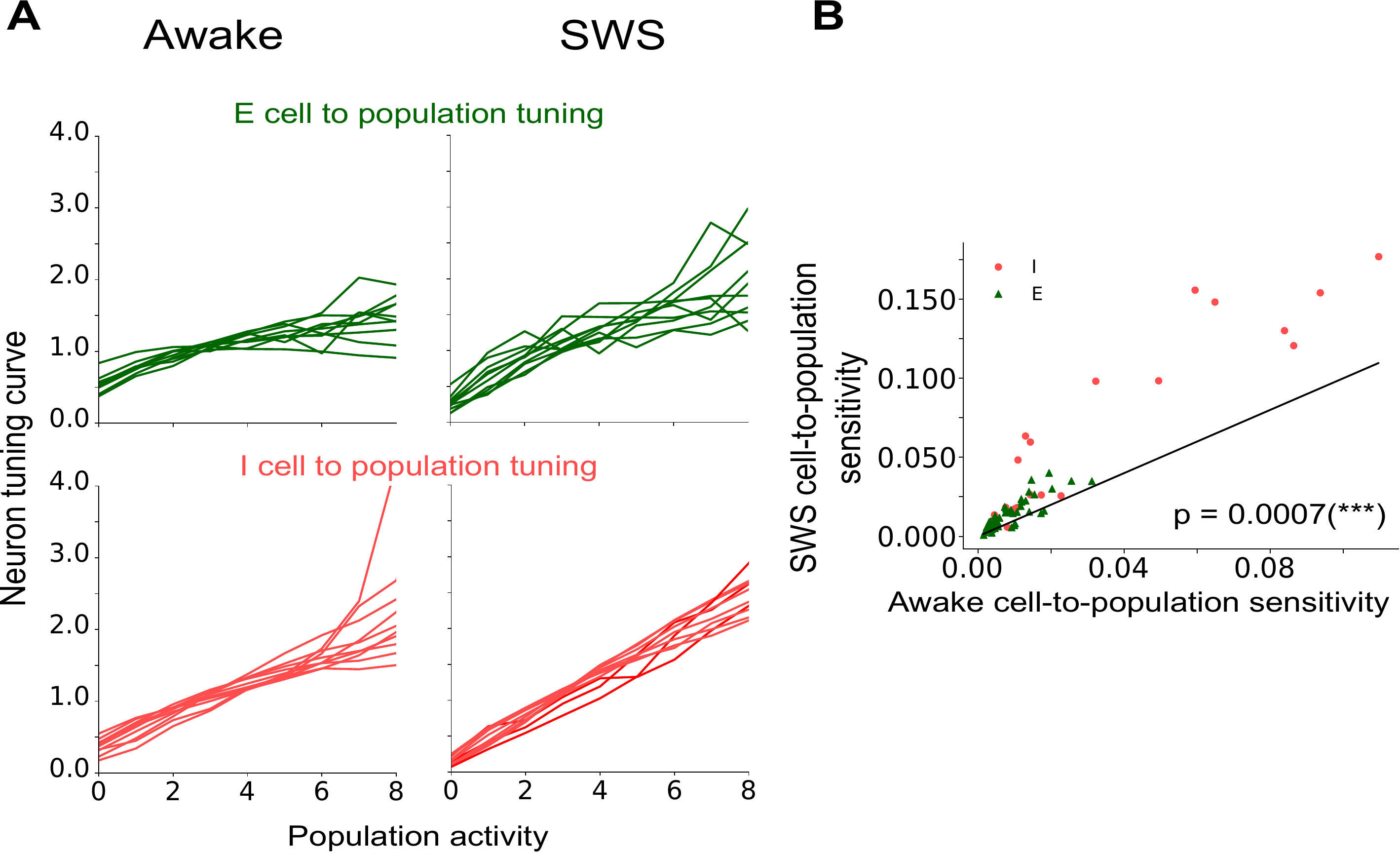}
\caption{
\textbf{Neural firing is tuned to the neural population's activity, particularly during SWS.} 
{\bf A}) Tuning curves of ten example neurons (see text and Appendix C) showing that neurons are tuned to \UF{the rest of the population's activity}.
{\bf B}) Scatter-plot  of the \TA{excitatory (green triangles) and inhibitory (red circles)} neuron sensitivity to the population activity (see Appendix C). 
Neurons are very consistently more sensitive during SWS ($p$-value $< 0.001$, Wilcoxon sing-ranked test).
}
\label{fig:tuning_all}
\end{figure*}

\begin{figure*}[t!]
\includegraphics[clip=true,keepaspectratio,width=1.99\columnwidth]{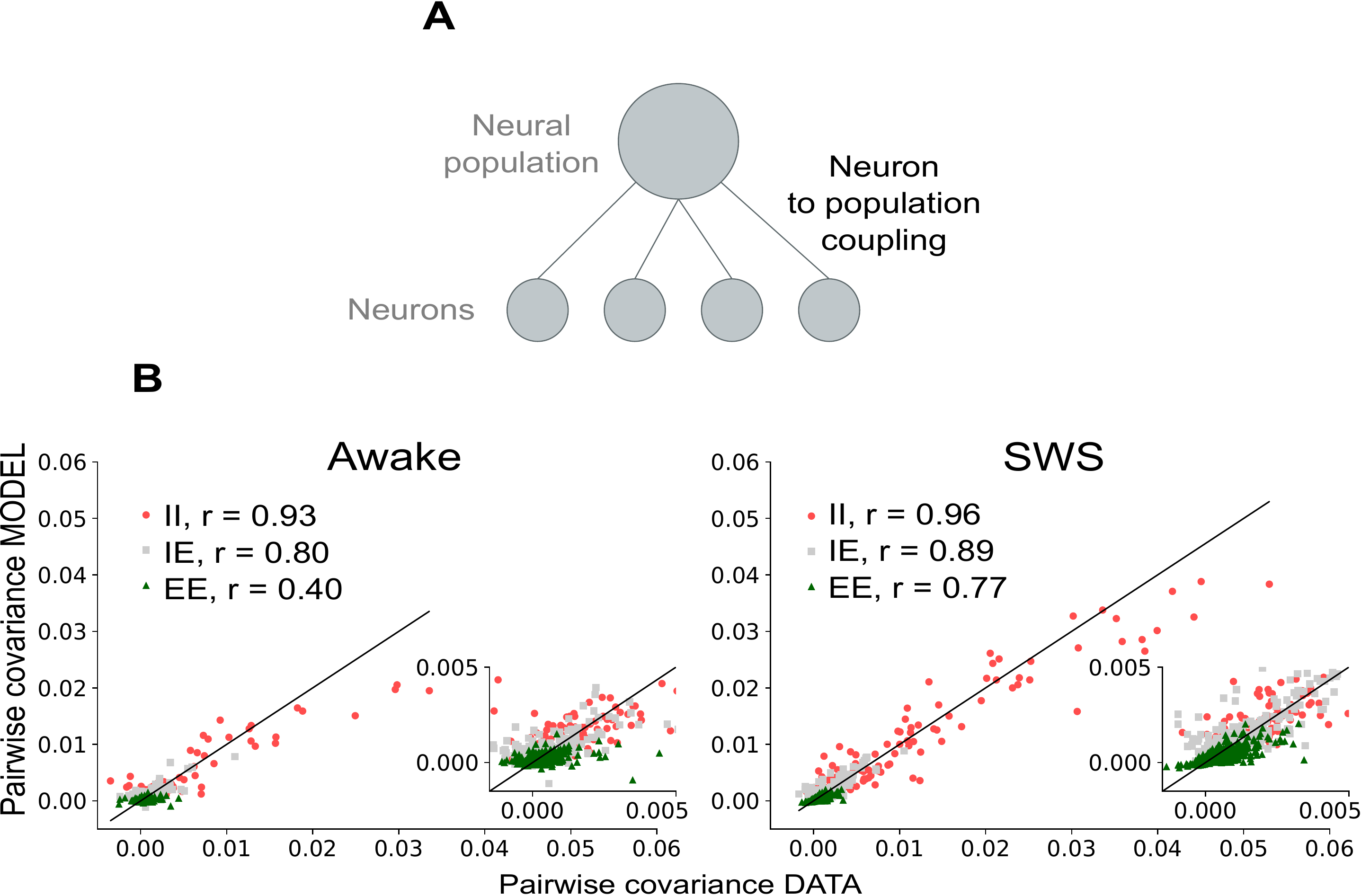}
\caption{\textbf{Single-population model shows better performance during SWS than wakefulness.}
{\bf A}) Model schematic diagram. 
{\bf B}) Pairwise covariances, empirical against predicted, for wakefulness (left) and SWS (right) states. 
Consistently with Fig.~\ref{fig:tuning_all}B, the success for SWS, most noticeably for I-I pairs \TA{(red circles)}, suggests these neurons are most responsive to whole-population activity.
\UF{Inset: enlargement of the small-correlation region.}
}
\label{fig:one-pop}
\end{figure*}

\begin{figure*}[t!]
\includegraphics[clip=true,keepaspectratio,width=1.99\columnwidth]{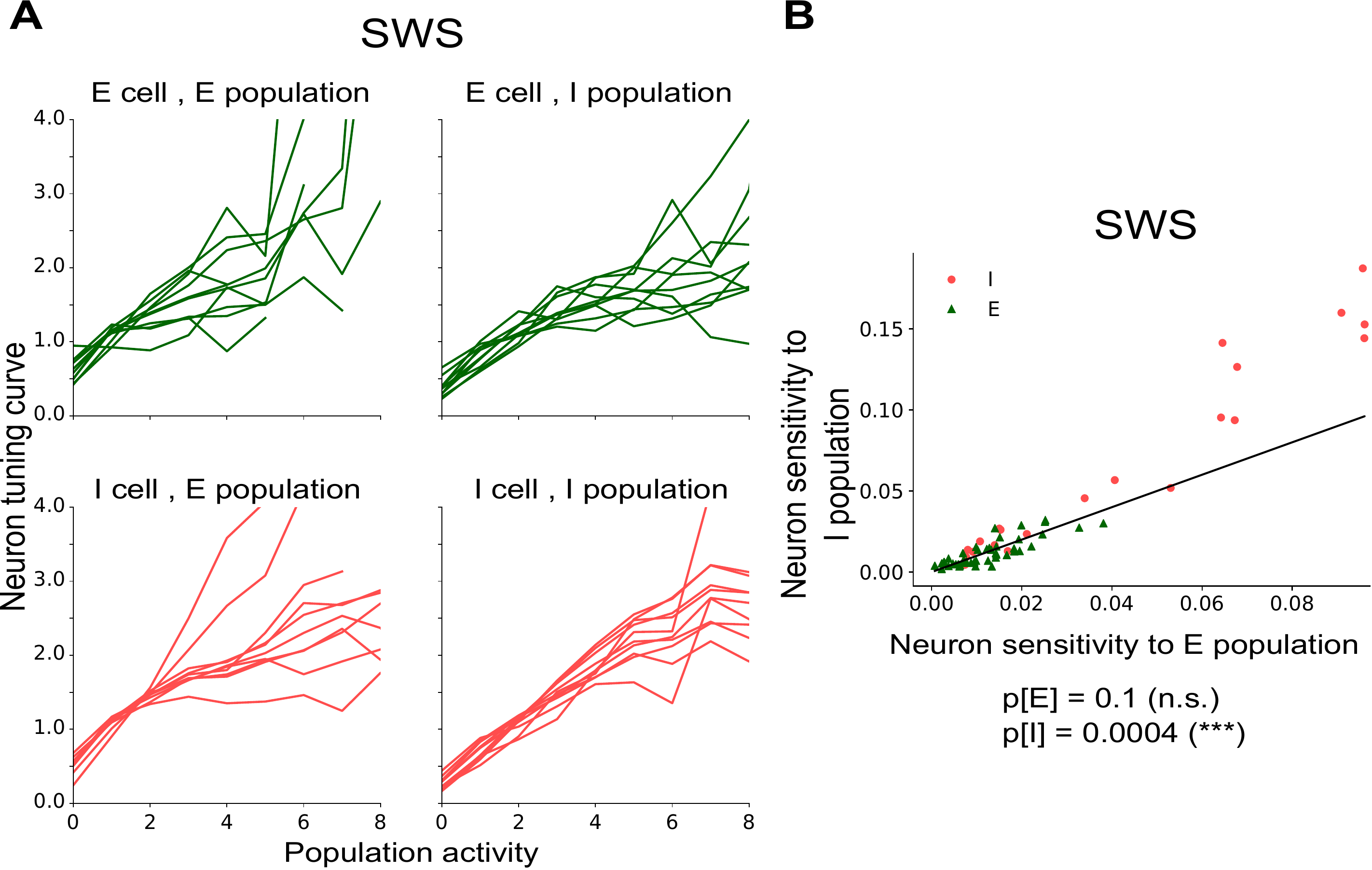}
\caption{\textbf{I neurons are more specifically tuned to the I population during SWS.} 
{\bf A}) Example tuning curves from ten neurons of each type to each type of population during SWS, and similarly for the I population. 
{\bf B}) Scatter-plot  of neuron sensitivity to E versus I population, during SWS.
I neuron \TA{(red circles)} are more tuned to I population than the E population ($p$-value $<10^{-3}$, Wilcoxon sign-ranked test).
E neurons \TA{(green triangles)}, instead, are weakly sensitive to both populations.
}
\label{fig:tuning_SWS_EI}
\end{figure*}

\begin{figure*}[ht!]
\includegraphics[clip=true,keepaspectratio,width=1.99\columnwidth]{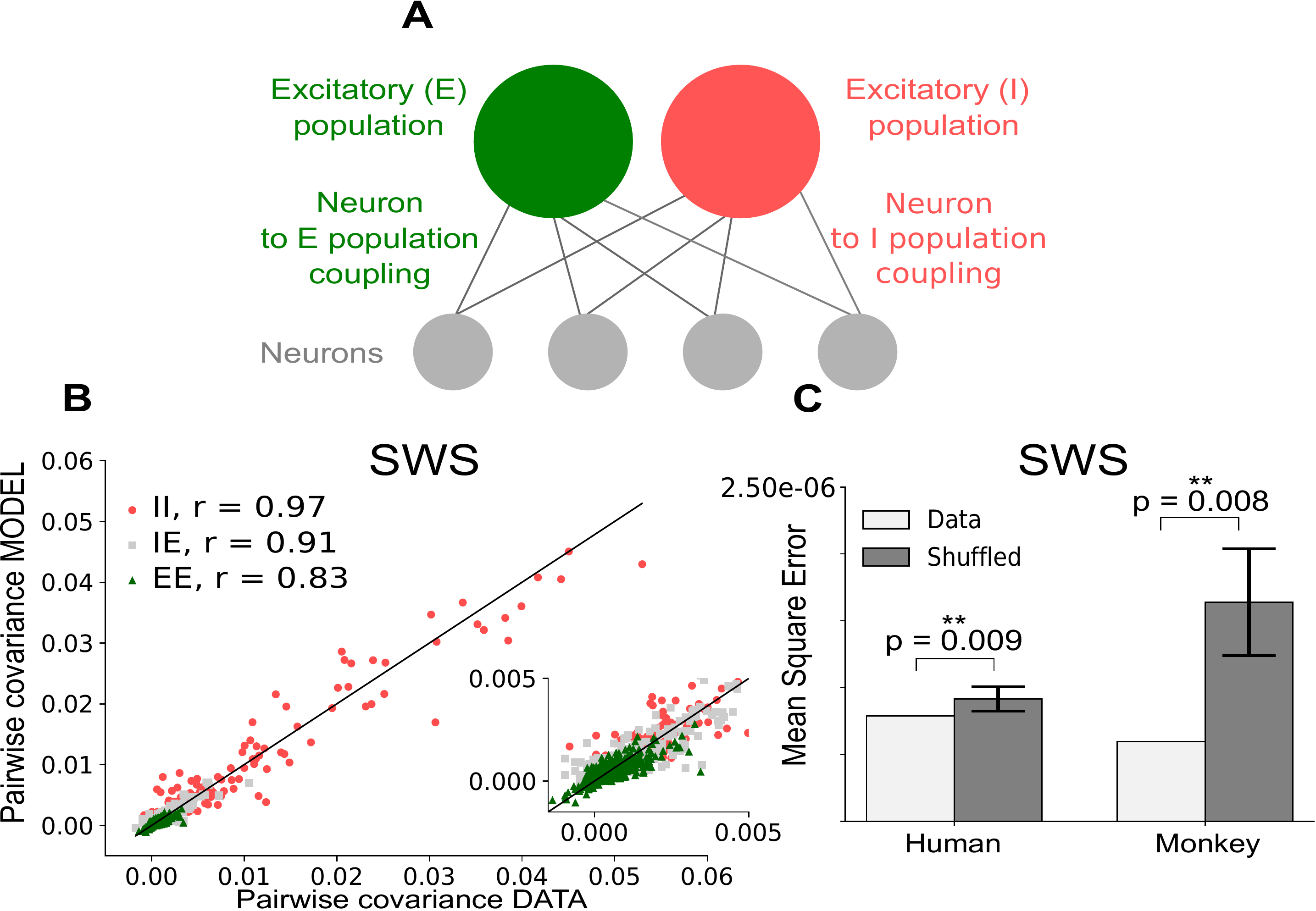}
\caption{
\textbf{Two-population model shows significant improvement in prediction for all types of neurons.}
{\bf A}) Schematic diagram of the two population model. 
Parameters $h^E_{iK^E}, h^I_{iK^I}$ are the couplings between each neuron $i$ and the E population activity $K^E = \sum_{i \in E} \sigma_i$ and the I population activity $K^I = \sum_{i \in I} \sigma_i$.
{\bf B}) Pairwise covariances, empirical against predicted, for the two-population model, during SWS. Improvement compared to the whole-population model is confirmed by the Pearson correlations. 
{\bf C}) Deterioration of prediction by shuffling neuron types for the human and monkey data-sets.
This effect demonstrates that knowledge of neuron types significantly contributes to improving model prediction. This is confirmed by the Mann-Whitney \textit{U} test p-values.
\UF{Inset: enlargement of the small-correlation region.}
}
\label{fig:two-pop}
\end{figure*}

\section{RESULTS}

We study 96-electrode recordings (Utah array) of spiking activity in the temporal cortex of a human patient and in the premotor cortex of a macaque monkey (see Appendix A), in wakefulness and slow-wave sleep (SWS), as shown in Fig. \ref{fig:raw_data}. Spike times of single neurons were discriminated and binned into time-bins of 50 ms (human data) and 25 ms (monkey data) to produce the population's spiking patterns (see Appendix A). From these patterns, we computed the empirical covariances between neurons then used for fitting models.

\subsection{Pairwise Ising model}
Pairwise correlations between I neurons have been found to exhibit invariance with distance \citep{Peyrache09}, even across brain regions \citep{LeVanQuyen16}. 
Here, we study what this intriguing observation implies for functional interactions between neurons, and the information conveyed by pairwise correlations on such interactions. 
Therefore, we investigate whether pairwise covariances are sufficient to capture the main features of neural activity, for E and I neurons during wakefulness and SWS.

To test this, we use a MaxEnt model that reproduces only and exactly the single neurons' spiking probability, and the pairwise covariances observed in the data.

As it has been shown \citep{Schneidman06,Cocco11b}, this model takes the form of a disordered Ising model (see Fig.~\ref{fig:Ising}A):
\begin{equation}
    P(\boldsymbol{\sigma}) = \frac{1}{Z}\exp\Big(\sum_i b_i\sigma_i + \sum_{i < j} J_{ij}\sigma_i \sigma_j \Big)
    \label{eq:psigma_ising}
\end{equation}
where $\sigma_i$ denotes activity of neuron $i$ given time bin (1: spike, 0: silence), $b_{i}$ the bias (or threshold) of neuron $i$, controlling its firing rate, and $J_{ij}$ the (symmetric) coupling between neurons $i$ and  $j$, controlling the pairwise covariance between the neurons.

We use the algorithm introduced by \cite{Ferrari16} to infer the model's parameters $b_{i}$ and $J_{ij}$ on data from wakefulness and SWS separately. Then we test how well the model describes neural activity in these states. 
In particular, synchronous events involving many neurons may not be well accounted for by the pairwise nature of the Ising model interactions.
To test this, we quantify the empirical probability of having $K$ neurons active in the same time window \cite{Tkacik14}: $K(\boldsymbol{\sigma}) \equiv \sum_i \sigma_i$. Fig.~\ref{fig:Ising}B compares the empirical probability distributions with model predictions. 
The Ising model is able to account for the empirical statistics during wakefulness, while it partially fails to capture the statistics during SWS.
This is confirmed by the measures of the Kullback-Leibler divergence, $D_\text{KL} \equiv \sum_K P_\text{data}(K) \log [P_\text{data}(K)/P_\text{model}(K)]$,  between empirical and model-predicted distributions (Fig.~\ref{fig:Ising}B). 
This difference can be ascribed to the presence of high activity transients, known to modulate neurons activity during SWS \citep{Steriade93} and responsible for the larger covariances, as seen in \cite{Peyrache12}.
In order to investigate the Ising model's failure during SWS, in Fig.~\ref{fig:Ising}C we compare the predictions for $P(K)$, separating \UF{E and I} neuron populations.
For periods of wakefulness, the model is able to reproduce both neuron types' behaviors.
However, during SWS periods, the model largely fails at predicting the empirical statistics, in particular for the I population.
This is confirmed by estimates of the Kullback-Leibler divergences (see Fig.~\ref{fig:Ising}).
Fig.~\ref{figSupp:Ising} shows similar results for the analysis on monkey recording.

These results highlight the relevance of the pairwise Ising model to reproduce $P(K)$ for all neurons, E and I, during wakefulness. Neural dynamics during wakefulness can therefore be described as predominantly driven by pairwise interactions. 
However, during SWS the model fails to reproduce $P(K)$ for both populations.
Therefore pairwise couplings alone are not sufficient and higher-order, perhaps even population-wide interactions may be needed to accurately depict neural activity during SWS. This is consistent with the observation that during SWS, neural firing is synchronous even across long distances, most notably for pairs of I neurons \citep{LeVanQuyen16}.

So far, our findings from inferring a pairwise Ising model on our datasets have highlighted that pairwise interactions were sufficient to depict neural activity during wakefulness, but higher-order, population-wide interactions may appear during SWS. 

\subsection{Single-population model}

In order to further characterize the neuronal activity during SWS, we consider the interaction between each neuron and the whole population: 
indeed, such approaches have proven successful in describing cortical neural activity \citep{Okun15}. 
We investigate whether neuron-to-population interactions exist in our data-set by studying the neurons' tuning curves to the population. 
Neuron-to-population tuning curves (see Appendix C) indicate how much a neuron's activity is determined by the total activity of the rest of the network \citep{Gardella16}.
In Fig.~\ref{fig:tuning_all}A we present tuning curves for ten example E or I neurons during both wakefulness and SWS.
These examples provide strong evidence for neuron-to-population tuning. 
In order to quantify population tuning, we estimate how much a neuron, either E or I, is \textit{sensitive} to the activity of the rest of the population, i.e. how much its activity fluctuates depending on the population activity (see Methods).
As can be observed in Fig.~\ref{fig:tuning_all}B, and consistently with our previous results, we find that neurons are sensitive to the population especially during SWS.
Similar results are valid for the monkey recording as well (Fig.~\ref{figSupp:one-pop}A).
Since we have established neuron-to-population interactions take place during SWS, we wish to determine to what extent they are sufficient in capturing the characteristics of neural activity during sleep.

To this purpose, we use a model \citep{Gardella16} for the dependencies between neuron firing, $\sigma_i=1$, and population activity, $k$: $P(\sigma_i = 1, k = K(\boldsymbol{\sigma}))$, where $K(\boldsymbol{\sigma})$ denotes the number of neurons spiking in any time bin. 
In this model (Fig.~\ref{fig:one-pop}A), the probability of neuron firing is described by the strength of its coupling to the population:
\begin{equation}
    P(\boldsymbol{\sigma}) = \frac{1}{Z}\exp\Big(\sum_i h_{ik}\delta_{k}^{K(\boldsymbol{\sigma})}\sigma_i\Big),
    \label{eq:psigma_delta}
\end{equation}
where $h_{ik}$ is the coupling between neuron $i$ and the whole population when $k$ neurons are active.
$\delta_k^K$ is the Kronecker delta, taking value one when the number $K$ of active neurons is equal to a given value $k$ and zero otherwise. 
For example, a ``chorister'' neuron, that fires most often when many others are firing, would have $h_{ik}$ increasing with $k$. Conversely, a ``soloist'' neuron, that fires more frequently when others are silent, would have $h_{ik}$ decreasing with $k$ \citep{Okun15}. $Z$ is the normalisation constant, that can be computed by summing over all possible activity configurations $Z = \sum_{\boldsymbol{\sigma}}\exp\Big(\sum_{i = 1}^{N} h_{ik}\delta_{k}^{K}\sigma_i\Big)$.  Importantly, $Z$ and its derivative allow us to determine the statistics of the model, such as the mean firing rate and the pairwise covariances. 
As an analytical expression exists for $Z$, the statistics may be derived analytically from the values of the couplings, making this model solvable (see Appendix \UF{D}).

To evaluate to what extent the model describes the data well and hence captures empirical statistics it was not designed to reproduce, we study the predicted pairwise correlations as compared to the empirical ones.

In Fig.~\ref{fig:one-pop}B, we compare the empirical pairwise covariances to their model predictions.
Pearson correlations (covariance between the two empirical and predicted variables, normalized the product of their standard deviations) confirm that the population statistics are better reproduced by the model during SWS than during wakefulness (Fig. \ref{fig:one-pop}). 
For monkey recording, the effect is even larger since the model entirely fails to account for wakefulness pairwise statistics (Fig.~\ref{figSupp:one-pop}B).
While the effect may be amplified by the fact that the Pearson correlations are larger during SWS, this is the opposite of what was observed for the pairwise Ising model: a model reproducing only empirical neuron-to-population interactions seems adequate at depicting neural dynamics during SWS but not during wakefulness. 

In particular, the model best reproduces the empirical statistics during SWS for \UF{I-I} neuron pairs. 
By contrast, E-E pairwise covariances are the most poorly reproduced during wakefulness. 
This result implies that during SWS, I activity, and to a lesser extent E activity, is dominated by population-wide interactions rather than local pairwise mechanisms, such that a MaxEnt 'population model' is mostly sufficient at capturing the key dynamics.
Nevertheless, this model still under-estimates the higher I-I pairwise covariances. 

\subsection{Two-population model}

Since I neurons are strongly synchronised even across long distances \citep{Peyrache12, Dehghani16}, we hypothesise that they could be tuned to the I population only, rather than the whole population.
We therefore ask if I neurons are tuned to the I population only. 
Indeed, as shown in Fig.~\ref{fig:tuning_SWS_EI}A, examination of the tuning curves of each neuron to the E and the I populations separately revealed homogeneous and strong tuning of I neurons to the I population, compared to tuning of I neurons to the E population or to the whole population (Fig.~\ref{fig:tuning_all}). 
In order to quantify this effect, we estimated the neuron sensitivity to both populations separately (see Appendix C).
The comparison in Fig.~\ref{fig:tuning_SWS_EI}B suggests I neurons are significantly more sensitive to the activity of I population than the E population. The effect is even larger for monkey recordings (Fig.~\ref{figSupp:two-pop}A).

To study tuning to the two populations separately, we now refine the previous model to take into account the couplings between each neuron and the E population and each neuron and the I population, separately. Because of the results of Fig.~\ref{fig:tuning_SWS_EI}B, we expect this model to perform better at reproducing the main features of the data during SWS. We want the model to only and exactly reproduce the empirical $P(\sigma_i = 1, k^E = K^E(\boldsymbol{\sigma}))$ and $P(\sigma_i = 1, k^I = K^I(\boldsymbol{\sigma}) )$ for all neurons $i$ and all values empirically taken by $K^E$ and $K^I$.

The probability of obtaining any firing pattern $\boldsymbol{\sigma}$ is given by (see Fig.~\ref{fig:two-pop}A)
\begin{equation}
    P(\boldsymbol{\sigma}) = \frac{1}{Z}\exp\Big(\sum_i (h_{ik^E}^E\delta_{k^E}^{K^E(\boldsymbol{\sigma})} + h_{ik^I}^I\delta_{k^I}^{K^I(\boldsymbol{\sigma})}) \sigma_i\Big),
    \label{eq:psigma_EI_delta}
\end{equation}
where $K^E(\boldsymbol{\sigma})$ is the number of E neurons spiking and $K^I$ the number of I neurons spiking in any time bin, and $h^E_{ik^E}$ the coupling between neuron $i$ and the whole E population when $k$ neurons are active, resp. $h^I_{ik^I}$ to the I population.
$Z$ the normalisation, $\delta_{k^E}^{K^E(\boldsymbol{\sigma})}$ and $\delta_{k^I}^{K^I(\boldsymbol{\sigma})}$ are Kronecker deltas as before. 
It can be shown (see Appendix D), following an analogous reasoning to that employed in \cite{Gardella16}, that this model is also analytically solvable in that the normalisation function $Z$ may be derived analytically. Using the expression for $Z$, as described in the Appendix \UF{D}, allows us to analytically predict the model statistics for any given set of couplings.
As for the previous models, we want to assess whether this model is sufficient to describe the data, that is if it can accurately predict a data statistic it was not specifically designed to reproduce. To this purpose we test pairwise covariances. We also aim to evaluate how prediction performance compares with the single-population model on the whole population (Fig. \ref{fig:one-pop}) described previously.

For both human (Fig.~\ref{fig:two-pop}B) and monkey (Fig.~\ref{figSupp:two-pop}B) recordings, during SWS the two-population model provides better predictions for pairwise covariances than the single-population model. 
Furthermore large I-I covariance are no longer systematically under-estimated. 
To verify the improvement in model performance was not solely due to this model possessing more parameters, we repeat the inference on the same data with the neuron types (E or I) shuffled, and find that the prediction deteriorates significantly, as highlighted in Fig.~\ref{fig:two-pop}C. 

\UF{A two-fold cross-validation test provided similar results for both data-sets, as the mean square error on the pairwise covariance prediction was smaller for the two-population model in the totality of trials (see Appendix D and Fig. \ref{figSupp:crossval}).}

\TA{Additionally, we note that the one-population model, Eq.~\ref{eq:psigma_delta}, inferred separately on the sub-populations of I neurons and E neurons, preforms similarly to the two-population model. This further supports that the knowledge of neuronal types is the key feature beyond the two-population model improvement (see Fig.~\ref{figSupp:sub-pop} for more details).} 

These \UF{analyses} demonstrate that taking into consideration each neuron's couplings with the E population and the I population separately is more relevant than taking into account its couplings with any sub-populations of the same size. \UF{We also note that while the deterioration due to shuffling is equally significant for both data-sets, it is more important for the monkey premotor cortex. This is consistent with the fact that E neurons, i.e. most neurons, are also \TA{very significantly} preferentially tuned to the I population for the monkey (Fig.\TA{~\ref{figSupp:two-pop}A}) but not for the human (Fig.\TA{~\ref{fig:tuning_SWS_EI}B}). Separating the two populations in the model therefore provides a much larger improvement on the prediction of E cells' behaviour in the monkey data. }

Remarkably, with the two-population model, E-I correlations are also reproduced with increased accuracy as compared to the single-population model. This improvement suggests that the two-population model successfully captures some of the cross-type interactions between the E and I populations, a non-trivial result since the two populations are not directly coupled to one another by design of the model.

\section{DISCUSSION}

In this paper, we tested MaxEnt models on human and monkey multi-electrode array recordings
where E and I populations were discriminated, during the states of
wakefulness and SWS. 
In order to investigate the properties of the neuronal dynamics, models were designed to reproduce one empirical feature at a time, and tested against remaining statistics.
 The pairwise Ising model's performance highlighted pairwise interactions as dominant in cortical activity during wakefulness, but insufficient to describe neural activity during SWS. 

We identify I neurons as responsible for
breaking pairwise sufficiency during SWS, suggesting instead that I
neurons' interactions are long-distance and population-wide, which
explains recent empirical observations \citep{Peyrache12,LeVanQuyen16}.

We found that models based on neuron-to-population
interactions, as introduced by \cite{Okun15}, are only relevant to
SWS, failing to replicate the empirical pairwise correlations
in the monkey premotor cortex (Fig.~\ref{figSupp:one-pop}). Even for SWS, I neurons' strong pairwise
correlations were consistently underestimated. 

Eventually, the two-population model provides a good trade-off for modelling neural interactions in SWS, and in particular the strongly correlated behaviour of I neurons. 
Discrimination between E and I neuron types greatly improves the capacity of a model to capture empirical neural dynamics.

\textbf{Pairwise sufficiency.} 
Pairwise Ising models (Fig.~\ref{fig:Ising}A) had previously been shown to accurately predict statistical features of neural interaction in many of data-sets \citep{Schneidman06,Cocco09,Hamilton13,Tavoni17}.
The surprisingly good performance of these models has raised hypotheses on the existence of some unknown mechanisms beyond their success \cite{Mastromatteo11}. 
In order to understand the so-called `pairwise sufficiency', a number of theoretical investigations \cite{Roudi09,Obuchi15a,Obuchi15b,Merchan16} and an empirical benchmark \cite{Ferrari17} have been conducted.
Model limitations have also been subject to some characterization. For instance, the breakdown of model performance for very large system sizes has been evidenced on experimental data \cite{Tkacik14} and studied theoretically \cite{Rostami17}. Ising model performance has also been shown to be sensitive to time bin size, and to its relation to characteristic time scales of the studied system \cite{Capone15}.
Here, we observed that for the same neural system, activity can be well-reproduced in one brain state (wakefulness) and not the other (SWS) (see Fig.~\ref{fig:Ising}B).
This result reinforces the idea that pairwise sufficiency depends on  the system's actual statistical properties, and it is not a more general consequence of the MaxEnt principle.

\textbf{\UF{Neuron}-to-population couplings}
 Although our study is the first to propose couplings between neurons and single-type population, an alternative approach has been previously used to highlight the neurons' tuning by the population activity \cite{Okun15}.
In that work, neurons were classified as `soloist' or `chorister', depending on whether they spiked more frequently when the rest of the population was silent or active, respectively.
\UF{Here}, we have refined this picture by pointing out tuning \UF{to a} single-type population.
Specifically, we have shown that I neurons are more sensitive to the I population activity than to the E one (Fig.~\ref{fig:tuning_SWS_EI}B). 
\UF{This result contributes to a literature having highlighted important synchrony between I neurons, including during sleep \cite{Peyrache12, LeVanQuyen16}. Our approach provides a complementary, quantitative view of this phenomenon in terms of neural interactions to the population. 
}

\UF{\textbf{Differences between data-sets and generality of results}
One should also note the different characteristics between the two data-sets we analyze. 
First, as seen in Fig.\ref{fig:raw_data}, neurons are less active
for the human data-set than the monkey. 
This difference may be due to recording in a different brain area \cite{Rolls90}, 
layer \cite{Sakata09}, and species \cite{Wallis12}. 
Second, neural correlations in the temporal and premotor cortex code for very different functions 
- long-term memory encoding in the temporal cortex \cite{Quiroga08}, and motion planning in the premotor \cite{Churchland10}.
While the differences above may justify any notable differences, 
namely the E neuron tuning to I population in SWS in the monkey data, 
it is important to highlight that all findings are consistent across both data-sets. 
This highlights that the framework we introduced is robust and may allow for further 
investigation of E and I dynamics and their interplay in a variety of empirical recordings.
Furthermore, this suggests the interactions uncovered here are not species or brain region-specific, 
but rather generic features of neural activity in the studied brain states. 
}

\UF{
\textbf{Competition between internal network dynamics and common external inputs.} 

We note that mechanisms underlying the neuronal interactions we observe can occur at multiple scales. Different network connectivity for I neurons \cite{Hofer11}, such as reinforced structural couplings over long distances, could account for the  population-wide interactions winning over pairwise interactions for I cells. Additionally, larger or more synchronous common inputs to the I population, across the scale of brain regions \cite{LeVanQuyen16, Olcese16}, may also be a plausible mechanism behind the observed interactions.
In conclusion, MaxEnt models can provide quantitative constraints to biophysical models of excitatory and inhibitory activity.
In turn, these biophysical models could serve the exploration of possible mechanisms 
 behind the observed neuron-to-neuron and neuron-to-population interactions.}

\section*{Acknowledgments}
We thank C. Capone, M. Chalk, M. di Volo, C. Gardella, J.S. Goldman, A. Peyrache, G. Tkacik and N. Tort-Colet for useful discussion.
Research funded by European Community (Human Brain Project, H2020-720270), ANR TRAJECTORY, ANR OPTIMA, 
French State program Investissements d'Avenir managed by the Agence Nationale de la Recherche [LIFESENSES: ANR-10-LABX- 65], 
NIH grant U01NS09050 and a AVIESAN-UNADEV grant.

\appendix
{
\section{DATA-SET}
We work with an intra-cranial multi-electrode array recording of 92 neurons in the temporal cortex of an epileptic patient, the same data-set used by \cite{Peyrache12} and \cite{Dehghani16}. The record of interest spans across approximately 12 hours, including periods of wakefulness as well as several stages of sleep. Recordings were performed in layer II/III of the middle temporal gyrus, in an epileptic patient (even though far from the epileptic focus and therefore not recording epileptic activity outside of generalised seizures). Data acquisition in that region was enabled by implanting a multi-electrode array, of dimensions 1 mm in thickness and 4x4 mm in area, with 96 micro-electrodes separated by 400 $\mu m$ spacings. The array was originally implanted for medical purposes. A 30-kHz sampling frequency was employed for recording. Switches in brain state (wakefulness, SWS, REM, seizure, ...) throughout the recording were noted from the patient's behavioural and physiological parameters, \UF{yielding three hours of wakefulness and one hour of SWS on which our analyses were focused}. Using spike sorting methods on the obtained data, 92 neurons have been identified. Analysis of the spike waveforms for each of these neurons allowed their classification as putative excitatory (E) and inhibitory (I) neurons. Using the spike times of each neuron, cross-correlograms for all pairs of neurons were also computed to determine whether each neuron's spikes had an excitatory (positive correlation) or an inhibitory (negative correlation) effect on other neurons through putative monosynaptic connections. It should be noted that neurons found to be excitatory exactly corresponded to those classified as RS, while all inhibitory neurons were also FS \UF{\citep{Peyrache12}}. We only retained neurons spiking all throughout the recording for our analyses, amounting to 71 neurons of which 21 were I neurons.\\
Similarly, spiking activity in layer III/IV of the premotor cortex of a macaque monkey was recorded by multi-electrode array, throughout a night, and an hour of target pursuit task on the following day. A 10-kHz sampling frequency was employed for recording. Classification of brain states was performed by visual inspection of the Local Field Potential (LFP), over time periods of 5 s, by identifying as SWS periods presenting large-amplitude oscillations in the 1-2 Hz frequency range. \UF{One hour of wakefulness and three hours of SWS data were used for our analyses}. Spike-sorting yielded 152 neurons, of which 141 spiked throughout the whole recording. Clustering on features of the spike waveform has allowed for the sorting of neurons as putative E and I \UF{\cite{Dehghani16}}. Excluding neurons for which clustering was uncertain within a 30-percent margin yielded 81 neurons, of which 38 were I, over which all subsequent analyses were performed as presented in \cite{Dehghani16}. \\
Time bin size was chosen in order to have one to few spikes from each neuron per time bin, while still having sufficient spikes per time bin from the whole population to compute statistics such as the pairwise covariances and the neuron-to-population dependencies. Since I neurons were consistently more active, this was equivalent to balancing a sufficient number of spikes from E neurons with sufficiently few spikes from any I neuron per time bin. In the human temporal cortex, where activity was considerably sparse, the chosen time bin size was 50 ms. In the interest of having comparable numbers of spikes per time bin and pairwise covariances, a time bin size of 25 ms was chosen for the monkey motor cortex, where firing rates were consistently higher than in the human temporal cortex (Fig. \UF{\ref{fig:raw_data}}). \\

\section{INFERENCE METHODS}
Inferring the parameters from a MaxEnt model may be understood as a Lagrange multiplier problem, where one maximises the entropy, while taking as constraints that the desired model-predicted statistics match their empirical values. Then each model parameter is the Lagrange multiplier for one constraint, on one observable to reproduce. Taking, for example, the pairwise Ising model, the statistics we want to reproduce are the neuron mean firing rates and the pairwise covariances. The corresponding model parameters are the firing thresholds $b_i$ and the pairwise couplings $J_{ij}$ respectively. We therefore want to maximise
\begin{align}
    S_{\mathrm{MaxEnt}} &= \max_{P} \min_{b, J} \left[ - \sum_{\boldsymbol{\sigma}}P(\boldsymbol{\sigma}) \log P(\boldsymbol{\sigma}) \right.& \nonumber \\
    &\left. + \sum_i b_i \Big(\sum_{\boldsymbol{\sigma'}} \sigma'_i  P(\boldsymbol{\sigma'})- <\sigma_i>_{\text{data}}\Big) \right.& \nonumber \\
    &\left. + \sum_{j\neq i} J_{ij} \Big(\sum_{\boldsymbol{\sigma'}} \sigma'_i\sigma'_j P(\boldsymbol{\sigma'}) - <\sigma_i\sigma_j>_{\text{data}}\Big)\right].&
   \label{eq:logL_ising}
\end{align}
One can verify that maximizing $S_{\mathrm{MaxEnt}}$ with respect to $P$ and with the chosen constraints \citep{Cocco11b,Gardella16}, gives the form of each of the models given previously in Eq. \ref{eq:psigma_ising}, Eq. \ref{eq:psigma_delta}, and Eq. \ref{eq:psigma_EI_delta}. 
A Hessian analysis may prove that the problem is well-posed, as the solution exists and it is unique.

The key challenge thus resides in finding a method that quickly converges to this solution.
Thanks to the explicit form of Eq.~(\ref{eq:logL_ising}), the gradient of the log-likelihood ($\ell  = -S_{\mathrm{MaxEnt}}$) with respect to the model parameters can be computed as differences between empirical and model-predicted averages of the conjugated observables.
For example, the gradient with respect to the bias $b_i$ of the Ising model can be estimated as $\langle \sigma_i \rangle_\text{data} -  \langle \sigma_i \rangle_\text{model}$.
The inference can thus be performed by an ascendant dynamics that  requires to estimate model averages of observables.
For the Ising model we applied the Markov-Chain Monte-Carlo method introduced in \cite{Ferrari16, Nghiem17}.
For the one population model, we applied the Newton dynamics proposed in \cite{Gardella16}
For the two-population model, we modify the algorithm of \cite{Gardella16} to take into account two populations. We found that a simpler steepest descent dynamics, that does not take into account the Hessian, was fast enough for our data-sets. 

\section{TUNING CURVE AND SENSITIVITY TO POPULATION}
In order to quantify the dependence of each neuron on the rest of the population's activity, we used tuning curves \citep{Gardella16} and sensitivity to the population.

\textbf{Tuning curves} characterize the dependence of the average activity of a neuron conditioned to the activity of either the population, either the E or I sub-population. 
The tuning curve of neuron $i$ is defined as $m_i(k)/\langle \sigma_i \rangle$, where $\langle \sigma_i \rangle$ gives the neuron's mean activity across all time bins. 
$m_i(k)$, instead, denotes the neuron's mean activity at fixed population activity and it is defined as: 
\begin{eqnarray}
m_i(k) &\equiv& P(\, \sigma_i =1 \,| \,K_{\setminus i}(\sigma) = k \,) \nonumber \\
&=& \frac{ P(\, \sigma_i=1\, , \, K_{\setminus i}(\sigma) = k \, )}{ P (\, K_{\setminus i} = k \, ) }
\end{eqnarray}
where $K_{\setminus i}(\sigma)$ is the number of active neurons  in the configuration $\sigma$, when neuron $i$ has been excluded. \UF{For example, $m_i(0)$ is the probability that neuron $i$ fires in time bins where all other neurons are silent.}

\textbf{Sensitivity to population.}
A tuning curve shows the whole profile of the dependence of a neuron activity on the rest of the population. 
In order to quantify this effect, we introduced the neuron \text{sensitivity} to the population, depicting the neuron's fluctuation in activity across states of population activity:
\begin{widetext}
\begin{equation}
\text{Sensitivity}_i \equiv \sqrt{\sum_k \Big(~ m_i^2(k) \, P (\, K_{\setminus i} = k \,)~\Big) - \sum_k \Big( ~m_i(k) \, P (\, K_{\setminus i} = k \, ) ~\Big)^2} = \sqrt{\sum_k \Big(~ m_i^2(k) \, P (\, K_{\setminus i} = k \, )~ \Big)  - \langle \sigma_i \rangle^2}~.
\end{equation}
\end{widetext}

\section{TWO-POPULATION MODEL}
In this section, we generalize the analysis of the one-population model introduced in \cite{Gardella16} to the case of two populations.
From our model introduced in Eq. \ref{eq:psigma_EI_delta}, we can define the couplings $h_{iK^E}^E \equiv h_{ik^E}^E\delta_{k^E}^{K^E}$ for E neurons to the E and the I populations, and respectively for I neurons, such that the probability of a firing pattern occurring is\\
\begin{equation}
    P(\boldsymbol{\sigma}) = \frac{1}{Z}\exp\Big(\sum_{i = 1}^{N} (h_{iK^E}^E + h_{iK^I}^I) \sigma_i\Big),
    \label{eq:psigma_EI}
\end{equation}
The model is said \textit{solvable} as the normalisation $Z$ can be expressed analytically. Note that the model is invariant under several gauge transformations as a number of linear combinations of its parameters $h_{iK^E}^E$ and $h_{iK^I}^I$ do not affect the probability distribution. 
Z and its derivative allow us to determine the statistics of the model, such as the mean firing rate and pairwise covariances.

\textbf{Normalisation.}
From Eq.~(\ref{eq:psigma_EI}) the normalisation is defined as
\begin{equation}
    Z = \sum_{\boldsymbol{\sigma}}\exp\Big(\sum_{i = 1}^{N} (h_{iK^E}^E + h_{iK^I}^I) \sigma_i\Big)
    \label{eq:Z_definition}
\end{equation}
where we sum over all possible firing patterns $\boldsymbol{\sigma}$. 
We may decompose this sum into terms $Z_{k^E,k^I}$ with given E and I population activities, such that
\begin{equation}
    Z =  \sum_{k^E = 0}^{N^E}\sum_{k^I = 0}^{N^I} Z_{k^E,k^I} 
    \label{eq:Z_kE_kI_definition}
\end{equation}
Then, we have
\begin{equation}
        Z_{k^E,k^I} = \sum_{\underset{K^E = k^E, K^I = k^I}{\boldsymbol{\sigma}}}\exp\Big(\sum_{i = 1}^{N} (h_{iK^E}^E + h_{iK^I}^I) \sigma_i\Big)\\
        \label{eq:Z_kE_kI_sum}
\end{equation}
where we sum over all possible firing patterns for all neurons for which $K^E$ excitatory neurons active and $K^I$ inhibitory neurons active. 
This is equivalent to summing over all possible patterns of E and I neurons independently, i.e.:
\begin{equation}
\begin{split}
    Z_{k^E,k^I} = \sum_{i^E_1 < ... < i^E_{k^E}}\sum_{i^I_1 < ... < i^I_{k^I}} \exp\Big(\sum_{b = 1}^{k^E} h_{i^E_bk^E}^E + h_{i^E_bk^I}^I\Big)\\\times\exp\Big(\sum_{c = 1}^{k^I} h_{j^I_ck^E}^E + h_{j^I_ck^I}^I\Big)\\
        = \Big[\sum_{i^E_1 < ... < i^E_{k^E}}\exp\Big(\sum_{b = 1}^{k^E} h_{i^E_bk^E}^E + h_{i^E_bk^I}^I\Big)\Big]\\ \times\Big[\sum_{i^I_1 < ... < i^I_{k^I}}\exp\Big(\sum_{c = 1}^{k^I} h_{i^I_ck^E}^E + h_{i^I_ck^I}^I\Big)\Big],
\end{split}
    \label{eq:Z_kE_kI_splitsum}
\end{equation}
where the $i^E_b$ spans over all the active E neurons for a given E activation pattern, and respectively for the $j^I_c$ for active I neurons. The result may be written as a product of two terms as these terms share no parameters in common. 

Here, the first term is summed over all possible firing patterns for E neurons that yield $K^E = k^E$, and similarly for I neurons in the second term. 
Now, analogously to \cite{Gardella16} let $Q$ be a polynomial such that the products over all i
\begin{equation}
    Q_E(X) = \prod_{\underset{i \in E}{i = 1}}^{N^E} 1 + X \exp(h_{ik^E}^E + h_{ik^I}^I),
    \label{eq:Q_EI_definition}
\end{equation}
where we take the product over all $i$ excitatory neurons, and similarly for $Q_I(X)$ multiplying over all inhibitory neurons. Now, the coefficient of $Q_E$ of order $X^{k^E}$, denoted $\mathrm{Coeff}[Q_E,X^{k^E}]$, corresponds to the sum over all the products of $k^E$ terms of the form $\exp(h_{ik^E}^E)$, or in other words, the sum over all products of combinations of E neurons $i^E_1 < ... < i^E_{k(E)}$, which is exactly equivalent to the first term of equation \ref{eq:Z_kE_kI_splitsum}. 
Since the same obviously applies for I neurons, we have
\begin{equation}
    Z_{k^E,k^I} = \mathrm{Coeff}[Q_E,X^{k^E}] \times \mathrm{Coeff}[Q_I,X^{k^I}],
    \label{eq:Z_kE_kI_coeff}
\end{equation}
As the $Q$ coefficients can be recursively computed, $Z$ is analytically computable, thus the model is solvable. 
Next we derive the statistics of the model from $Z$.

\textbf{Model statistics.}
The statistics predicted by the model is given by differentiating $Z$. We use this to predict the mean firing rates and pairwise covariances from the population couplings $h_{ik^E}^E$ and $h_{ik^I}^I$. 

As we defined in Eq.s~(\ref{eq:Z_definition}) and (\ref{eq:Z_kE_kI_definition}),
\begin{equation}
    Z = \sum_{k^E = 0}^{N^E}\sum_{k^I = 0}^{N^I}\exp\Big(\sum_{i = 1}^{N} (h_{ik^E}^E\delta_{k^E}^{K^E} + h_{ik^I}^I\delta_{k^I}^{K^I}) \sigma_i\Big), 
    \label{eq:Z_sum_delta}
\end{equation}
Thus, the joint probability of a given neuron spiking and a given number of E neurons spiking in any time bin is as follows:
\begin{equation}
    P(\sigma_i = 1, k^E = K^E) = <\sigma_i \delta_{k^E}^{K^E}> = \frac{\partial ln Z}{\partial h_{iK^E}^E}
    \label{eq:mean_sigma_model_partial}
\end{equation}
Recalling our expression for $Z_{k^E,k^I}$ in Eq.~(\ref{eq:Z_kE_kI_coeff}), this yields
\begin{eqnarray}
&&P(\sigma_i = 1, k^E = K^E) = \nonumber \\
&& ~~\frac{1}{Z}\sum_{k^I}\Big(\mathrm{Coeff}[Q'_E,X^{K^E}] \mathrm{Coeff}[Q_I,X^{k^I}]\Big) \label{eq:mean_sigma_model_coeff}
\end{eqnarray}
where 
\begin{eqnarray}
Q'_E &\equiv&\frac{\partial ln Z}{\partial h_{iK^E}^E} \mathrm{Coeff}[Q_E,X^{k^E}] \nonumber \\
&=& X e^{h_{iK^E}^E + h_{ik^I}^I}\prod_{j\neq i}(1+X e^{h_{jK^E}^E + h_{jk^I}^I})
\end{eqnarray}

and $i$ is an E neuron. For an I neuron's firing probability one can swap around E and I in Eq.~(\ref{eq:mean_sigma_model_coeff}). 
This allows the straightforward derivation of the firing rate by summing over all values of $K^E$ (resp. $K^I$ for I neurons).

Likewise, the pairwise correlations may be computed from
\begin{equation}
    <\sigma_{i_E}\sigma_{j_I}> = \frac{1}{Z}\sum_{k^E}\sum_{k^I}\Big(\mathrm{Coeff}[Q'_E, X^{k^E}]\times\mathrm{Coeff}[Q'_I, X^{k^I}]\Big)
    \label{eq:pairwise_corr_model_difftypes}
\end{equation}
for two neurons of different types, and
\begin{equation}
    <\sigma_{i_E}\sigma_{j_E}> = \frac{1}{Z}\sum_{k^E}\sum_{k^I}\Big(\mathrm{Coeff}[Q''_E, X^{k^E}]\times\mathrm{Coeff}[Q_I, X^{k^I}]\Big)
    \label{eq:pairwise_corr_model_sametypes}
\end{equation}
where 
\begin{equation}
    Q''_E = X^2 e^{h_{ik^E}^E + h_{ik^I}^I+h_{jk^E}^E + h_{jk^I}^I}\prod_{l\neq i,j}(1+X e^{h_{lk^E}^E + h_{lk^I}^I})
\end{equation} and i is an E neuron (extending to I neurons is very straightforward).

\textbf{Shuffle tests.} To verify whether information on neuron types significantly improves the model's prediction performance, for each species, we perform a series of ten inferences on the same SWS data-set. Each time, the neuron labels are independently shuffled, while the number of E neurons and the number of I neurons remains the same. The Mean Square Error (MSE) on the predicted pairwise covariances is computed every time. We found it to be consistently larger for the shuffled trials compared to that where empirical neuron types are known. This is quantified by the Mann-Whitney \textit{U} test on the samples of $\text{MSE}_{\text{shuffled}}[n] - \text{MSE}_{\text{data}}$ with n ranging from one to ten.\\

\UF{\textbf{Cross-validation.} As a means to further verify the robust improvement of the two-population model over the single-population model, we perform a two-fold cross-validation. Half of the time bins are chosen at random, on which the single-population and two-population models are inferred. The pairwise covariances predicted by these inferred models are compared against the empirical pairwise covariances computed on the other half of the time bins, to obtain the MSE for each of the two models. This process is repeated $15$ times on each data-set; in all of the repetitions the MSE is smaller for the two-population model. The improvement is thus statistically significant, as confirmed by the Wilcoxon signed-rank test ($p$-value $<10^{-3}$, Fig. \ref{figSupp:crossval} \cite{SupplMat}).} 
}

\section*{Supplementary figures}
 \renewcommand{\thefigure}{S\arabic{figure}}
\setcounter{figure}{0}    

\begin{figure*}[h!]
\includegraphics[clip=true,keepaspectratio,width=1.8\columnwidth]{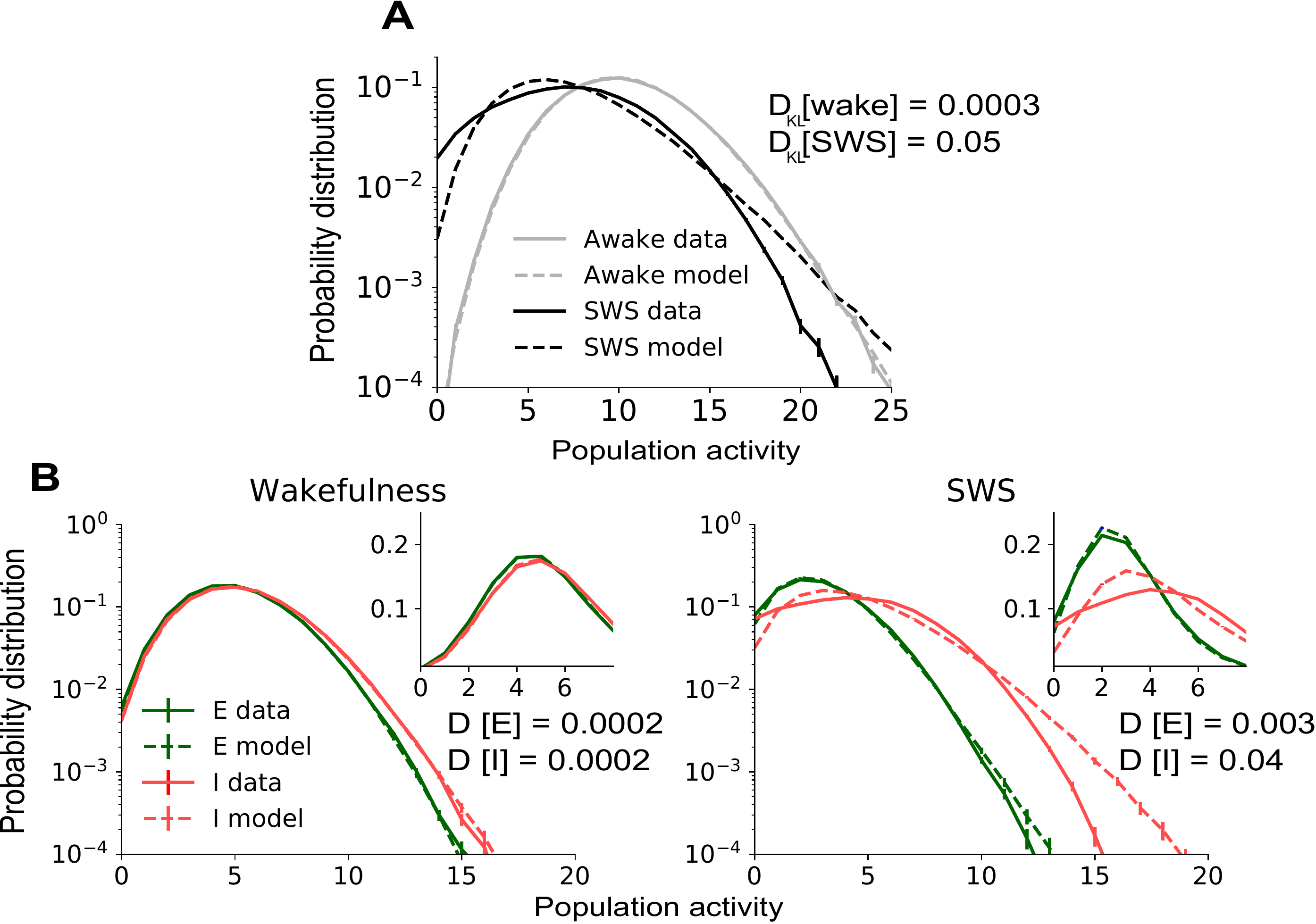}
\caption{
\textbf{Pairwise Ising model analysis on monkey recording.}
{\bf A}) Empirical and predicted distributions of the population activity $K$ for the population of excitatory (E) neurons, and that of inhibitory (I) neurons. 
On the monkey data, the Ising also performs better at capturing the population statistics during wakefulness than SWS. 
{\bf B}) Empirical and predicted population activities for E and I neurons. The model fails at reproducing the statistics of inhibitory population activity during SWS, similarly to with the human data. \UF{Insets show an enlarged view on the region of low population activity in linear scale}.
}
\label{figSupp:Ising}
\end{figure*}

\begin{figure*}[h!]
\includegraphics[clip=true,keepaspectratio,width=1.8\columnwidth]{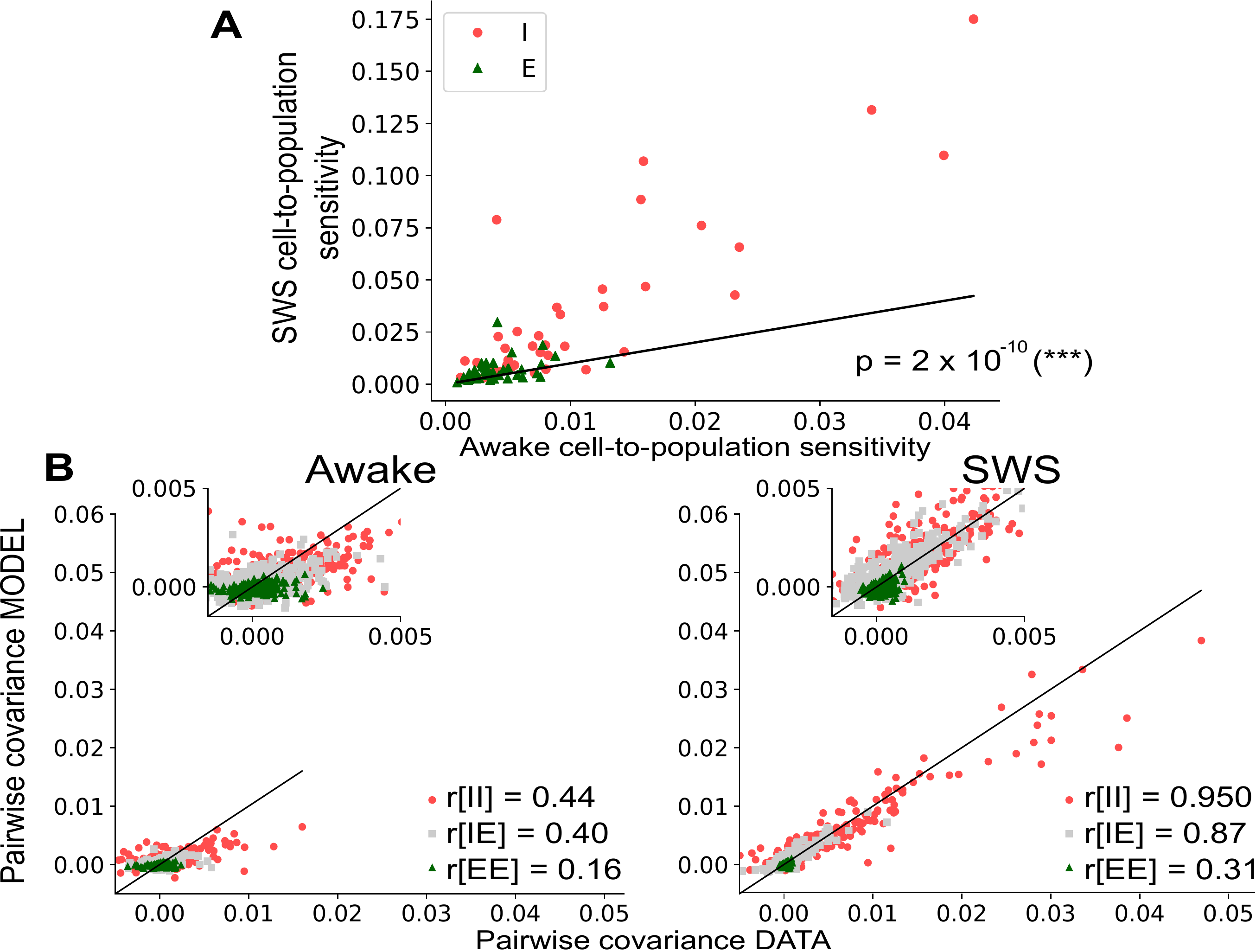}
\caption{\textbf{Single-population model analysis on monkey recording.}
{\bf A}) Scatter-plot of the neuron sensitivity to the population activity in wakefulness and SWS. 
Neurons are consistently more sensitive during SWS ($p$-value $<0.001$).
{\bf B}) Pairwise covariances, empirical against predicted, for wakefulness (left) and SWS (right) states. 
Relative success for SWS, especially I-I pairs suggests these neurons are most responsive to whole-population activity, even though the model tends to under-estimate the larger pairwise covariances.
The model completely fails to account for pairwise covariances during wakefulness.\UF{ Inset: enlargement of the small-correlation region.}
}
\label{figSupp:one-pop}
\end{figure*}

\begin{figure*}[h!]
\includegraphics[clip=true,keepaspectratio,width=1.8\columnwidth]{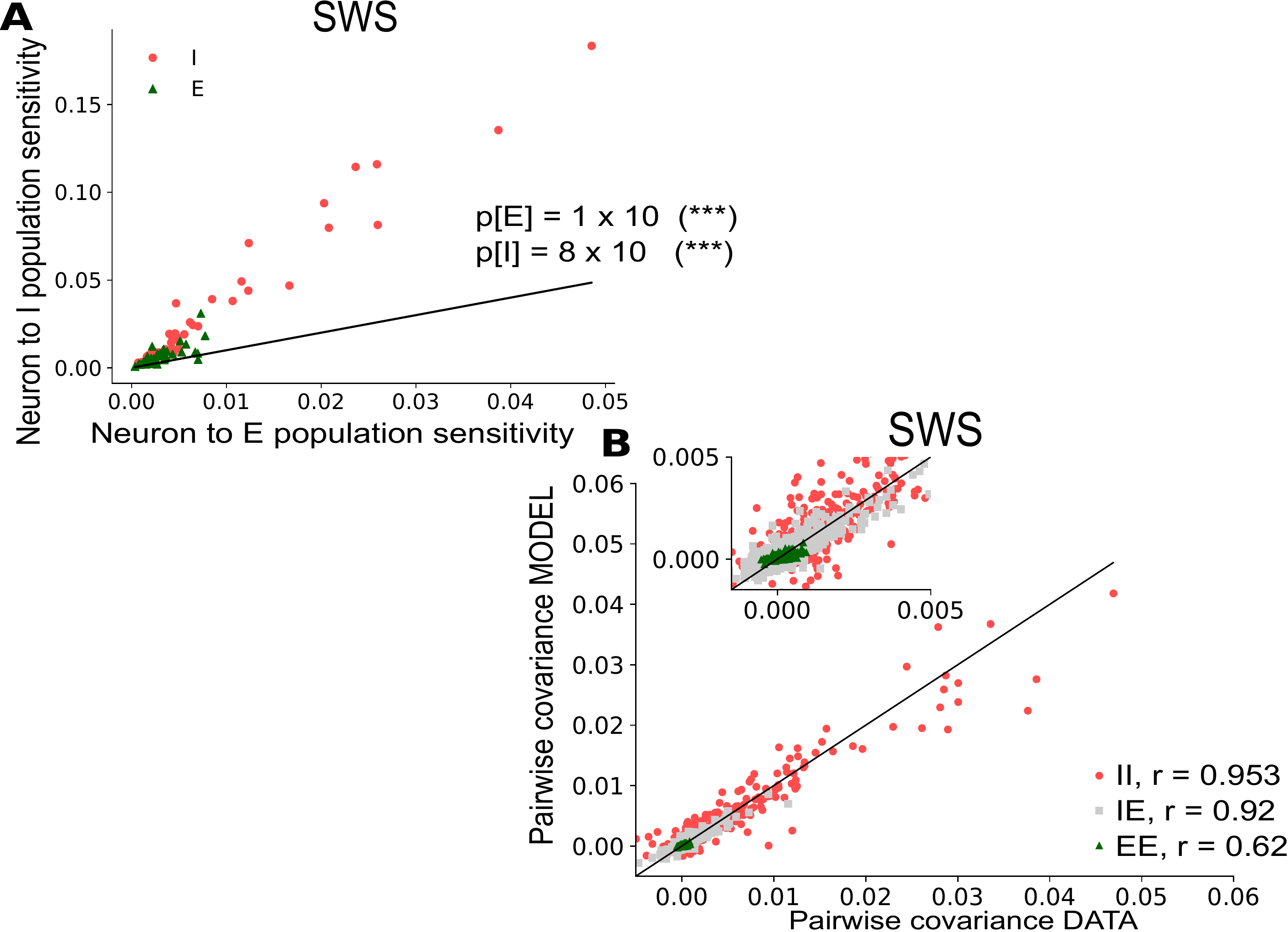}
\caption{\textbf{Two-population model shows significant improvement in prediction for all types of neurons.}
{\bf A}) Scatter-plot of neuron sensitivity to E versus I population, during SWS.
Both I and E neurons are more tuned to I population than the E one ($p$-value $<10^{-3}$, Wilcoxon sign-ranked test).
{\bf B}) Pairwise covariances, empirical against predicted, for the two-population model, during SWS. 
Improvement compared to the whole-population model is confirmed by the Pearson correlations. \UF{Inset: enlargement of the small-correlation region.}
}
\label{figSupp:two-pop}
\end{figure*}

\begin{figure*}[h!]
\includegraphics[clip=true,keepaspectratio,width=1.8\columnwidth]{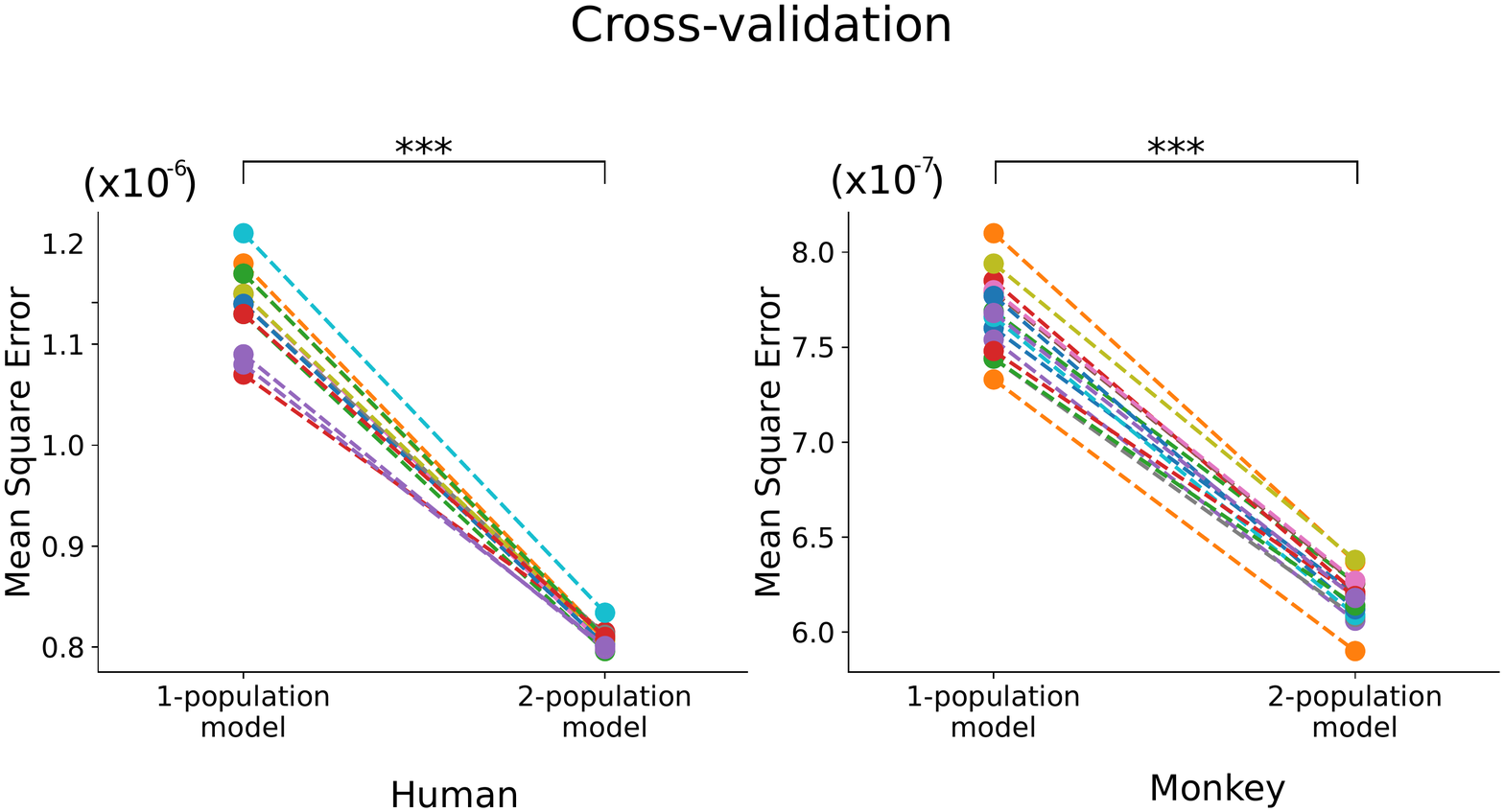}
\caption{\UF{\textbf{Two-population model outperforms single-population model at estimating pairwise covariances, as shown by cross-validation.}
Mean Square Error (MSE) between model-predicted pairwise covariances (where the model is inferred a randomly chosen half of the data) and their empirical counterparts (from the remaining half, see Appendix D).
For all 15 trials (different in colors), the MSE is always smaller for the two-population model than the one-population model ($p$-value $<10^{-3}$, Wilcoxon sign-ranked test).}
}
\label{figSupp:crossval}
\end{figure*}

\begin{figure*}[h!]
\includegraphics[clip=true,keepaspectratio,width=1.8\columnwidth]{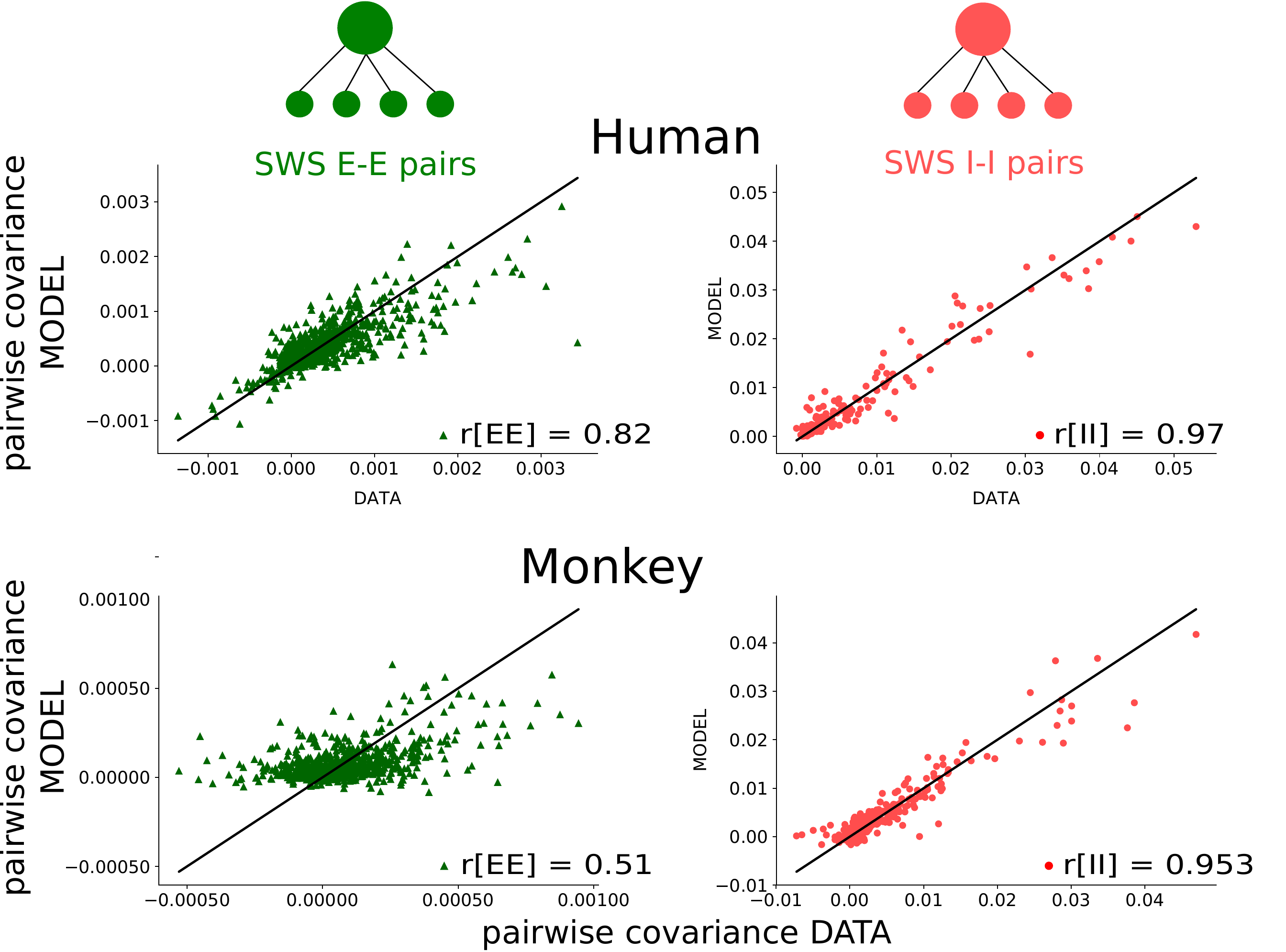}
\caption{\TA{\textbf{One-population model inferred on E and I sub-populations separately can account for most of the improvement of the two-population model, especially for inhibitory neurons.}
For both human and monkey data-sets, the one-population model is inferred on the sub-population of E neurons (left), and on that of I neurons (right). Comparing the empirical covariances with their model-predicted counterparts, the model performance is seen to be very similar to the two-population model's. Indeed, this 'sub-population' model accounts for the majority of the improvement of the two-population model over the one-population model inferred over the whole population. The fraction of improvement, $(r_{subpop} - r_{1-pop})/(r_{2-pop} - r_{1-pop})$, is larger for I neurons (human: 89\% ; monkey: 95\%) than for excitatory neurons (human: 83\%; monkey: 65\%). This is consistent with a strong tuning of I cells to the I population (Fig.~\ref{fig:tuning_SWS_EI}).}
}
\label{figSupp:sub-pop}
\end{figure*}

\end{document}